\documentclass[aps,prb,twocolumn,superscriptaddress,longbibliography,floatfix]{revtex4-2}

\usepackage{graphicx}
\usepackage{dblfloatfix}
\usepackage{amsmath, amssymb}
\usepackage{xcolor}
\usepackage{hyperref}
\usepackage{physics}

\begin{document}

\title{Thermodynamic Driving Force Activated Phonon Scattering in InN}

\author{Zaheer Ahmad}
\affiliation{Department of Physics, University of Illinois, Chicago, IL 60607, USA}

\author{Osama A. Rana}
\affiliation{Department of Electro-Optics and Photonics, University of Dayton, Dayton, OH 45469, USA}

\author{Shakeel Ahmad}
\affiliation{Department of Physics, University of Alabama, Birmingham, AL 35233, USA}

\author{Mark Vernon}
\affiliation{Joseph Maxwell Cleland Atlanta VA Medical Center, Decatur, GA 30033, USA}

\author{Brendan Cross}
\affiliation{Department of Physics, Andrews University, Berrien Springs, MI 49104, USA}

\author{Alexander Kozhanov}
\affiliation{Department of Electrical and Computer Engineering, Duke University, Durham, NC 27705, USA}

\date{\today}

\begin{abstract}
Defect related disorder during InN growth is a major challenge for making high performance electronic and optoelectronic devices. This is partly because film quality is often described using reactor specific settings instead of general physical variables. In this study, we show that plasma assisted MOCVD growth of InN can be described using a single thermodynamic driving force coordinate. This coordinate brings together growth kinetics, defect sensitive Raman response and structural coherence across different process conditions. When we use this coordinate, the incorporation rate follows a universal activated trend with a kinetic scale of about 0.08 eV. Raman measurements show a clear crossover between a defect sparse and a defect rich regime: a disorder activated Raman metric increases quickly after the crossover, while an A1–LO control metric stays mostly the same. This suggests that short range lattice disorder, not long range polar coupling, dominates the defect activation process. X-ray diffraction shows that the out of plane coherence length stays the same for samples with the same driving force, even if reactor settings are very different. This supports the idea that structural coherence is organized by thermodynamics in this growth window. Finally, a simple kinetic Monte Carlo model using driving force biased incorporation and defect activation events matches the observed exponential trends and the two regimes, supporting the driving force approach. These results show that a transferable driving force coordinate can be used for plasma assisted InN growth and offer a quantitative way to achieve defect sparse growth conditions.
\end{abstract}

\maketitle

\section{Introduction}

Group III nitride semiconductors are behind a large part of modern high power\cite{Dobrinsky2013ECS,Zhang2021RPP,Khan2005PESC,Zhou2017ECS} and high frequency\cite{Hao2012JOS,Giannazzo2017PSSA,Asbeck2019SemSem,Simin2013SST} electronics, solid state lighting\cite{Wierer2015PSSA,Krames2002PSSA,Wierer2014PSSC,Wierer2016LPR,Shen2019JSTQE} and many sensing platforms\cite{Khan2005JJAP,Schalwig2002MSEB,Schalwig2001PSSA,Pau2006ApplOpt,Li2017Nanoscale}. In most cases the performance limit comes not from the ideal band structure that we calculate but from the disorder created during processing\cite{Zolper1997JCG,Ahmad2025APLNiImplant,Calleja2007PSSB,Ahmed2023arXivNi} including the growth\cite{Minj2015ActaMat,Tuomisto2012JCGVacancy,Paskova2010ProcIEEE,Uedono2009JCGPositron,SmalcKoziorowska2020SciRep} itself. In the nitrides, defects are not a small correction. They set the carrier density, the dominant scattering processes, the reliability of the device over time and even how long phonons stay coherent.
\par
Most growth reports describe the process in terms of settings on the reactor such as precursor flow ratios, chamber pressure, plasma power and substrate temperature. Each of these settings changes several microscopic processes at once. The same written recipe can therefore correspond to very different local chemical conditions in different reactors or even in the same reactor at different times. Empirical rules that seem to work in one tool can fail completely in another.
\par
Indium nitride is a natural material for testing whether a similarly grounded description can be built for growth. The narrow band gap, strong ionicity and poor thermal stability make adsorption, incorporation, desorption and defect formation compete strongly at the growth surface. In plasma assisted metalorganic chemical vapor deposition the supply of active nitrogen is set by the plasma rather than by simple thermal chemistry. This opens growth regimes that conventional thermal MOCVD cannot reach. At the same time the near surface environment is strongly nonequilibrium and chemically crowded. Small changes in precursor supply or plasma conditions can drive large changes in the final defect landscape. This leads to a simple question. Instead of working in the full space of reactor settings can we reorganize the problem in terms of a small number of variables that actually track growth rate, disorder signatures and structural coherence.
\par
One obvious candidate is the thermodynamic driving force for incorporation. In many epitaxial systems differences in chemical potential give a compact way to connect growth rate, nucleation behaviour and defect energetics. For plasma assisted InN growth this idea is harder to use directly because the activity of reactive nitrogen cannot be measured and depends on plasma conditions\cite{Ahmad2019APLInN} in ways that equilibrium gas phase models do not capture. The basic picture is still useful but it needs an operational surrogate. Here we construct such a surrogate in the form of a driving force axis that is thermodynamically consistent and that can be evaluated from the process conditions. We then use this single coordinate as an organising variable for the experimental data. The spirit is similar to other physics informed modelling work in many domains\cite{Yang2019PRF,Ahmad2025ResearchSquareConfocal,Sharma2023Energies,Champion2020IEEEAccess,Ahmad2025arXivAutoencoders,Karniadakis2021NatRevPhys,Ahmad2026SciRepNoise}, where complicated measurement chains are rewritten in terms of a few coordinates tied to real physical mechanisms rather than to raw knob settings.

\par
In the present study we build and test this effective driving force framework for plasma assisted MOCVD of InN building on our previous efforts\cite{Ahmad2023arXivPlasmaPotential,Ahmad2020arXivSupersaturation}. Instead of treating each reactor setting separately we track one scalar quantity that represents the net bias for incorporation under the combined influence of precursor supply, total pressure and plasma conditions. We then examine how this single coordinate orders three kinds of observable. These are the macroscopic growth rate, Raman features that respond to disorder and x ray diffraction measures of coherence for samples grown under quite different nominal conditions. To connect these trends to a microscopic picture without attempting a full plasma chemistry model we also construct a minimal kinetic Monte Carlo simulation in which a small set of activated surface events are biased by the same driving force parameter. The combination of experiment and simulation then lets us test whether the behaviour of the system can be understood in terms of a simple and physically interpretable mechanism rather than an opaque list of reactor settings.

\par
All layers were grown at 775C\cite{Cross2020JCGInN} and activated nitrogen concentration was calculated from in-situ plasma emission spectra following the method described in our previous studies\cite{Ahmad2019APLInN}.

\section{EXPERIMENTAL METHODS}
InN films were grown on $c$ plane Al$_2$O$_3$ substrates in a vertical flow plasma assisted MOCVD reactor with an inductively coupled RF nitrogen plasma source operating at 13.56~MHz. Before growth the substrates were thermally outgassed inside the reactor. Growth started with a thin low temperature InN nucleation layer about 2 to 3~nm thick at $500~^\circ\mathrm{C}$ to improve wetting and to suppress patchy initial coverage. After this step the temperature was raised to 1048~K which is $775~^\circ\mathrm{C}$ and kept fixed for all samples in this study. Only the gas phase conditions were changed from run to run. The TMIn molar flow was varied between 0.8 and 9~$\mu\mathrm{mol}\,\mathrm{min}^{-1}$. The reactor pressure was set between 2.2 and 4.0~Torr. RF plasma power was set between 150 and 550~W. The N$_2$ carrier flow was kept between 0.6 and 1.0~standard liters per minute and was not used as an independent control variable.

To reduce this many parameter space to a single control coordinate we define a dimensionless supersaturation index $\sigma$ from the experimentally controlled process conditions. In practice $\sigma$ is constructed to increase monotonically with the effective indium chemical activity at fixed temperature. Here this is done by referencing the supplied TMIn molar flux to a low flow baseline and then applying a correction factor that accounts for plasma activation normalized by pressure. From $\sigma$ we define an effective driving force coordinate
\begin{equation}
\Delta\mu = k_{\mathrm{B}} T \ln\!\left(1 + \sigma\right)
\end{equation}

Throughout this work $\Delta\mu$ is used as an experimentally operational driving force axis. It is not intended to represent the literal chemical potential difference of any single identified gas phase species in the plasma.

Film thickness was measured optically from FTIR reflection spectra and growth rates were obtained by dividing the thickness by the deposition time. Structural characterization was carried out using high resolution x ray diffraction on a PANalytical system. The 0002 reflection was used to extract the out of plane coherence length $D_{0002}$ using the Scherrer relation. For the subset of samples at low $\Delta\mu$ the 0002 and 0004 reflections were also used together in a Williamson Hall analysis to estimate the microstrain $\varepsilon$ and the $c$ axis interplanar spacing $d_{0002}$.

Raman spectra were collected at room temperature using 532~nm excitation in backscattering geometry with an optical spot size of about 1~micrometer. All spectra were acquired with the same optical alignment and excitation power density across samples in order to minimize systematic intensity differences. Individual peaks were fit with Lorentzian line shapes to obtain the linewidths

\begin{figure*}[t]
    \centering
    \includegraphics[width=0.99\textwidth]{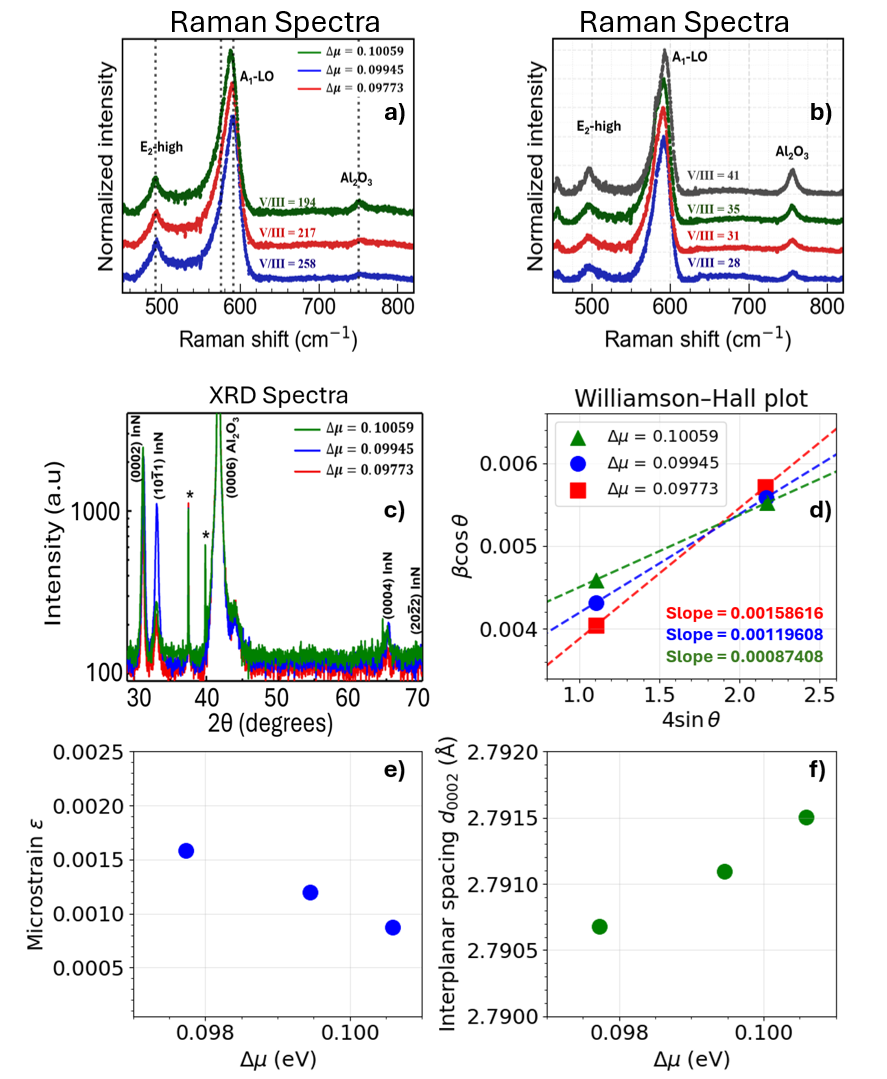}
    \\
    {\textbf{Figure 1:} Low-$\Delta\mu$ baseline linking spectroscopy and structure. Representative normalized Raman spectra show the $E_2^{\mathrm{high}}$ and $A_1$–LO modes with sapphire ($\mathrm{Al_2O_3}$) features retained for reference (a,b). Corresponding $\theta$–$2\theta$ scans (log intensity) show the InN (0002) and (0004) reflections alongside substrate peaks (c). Williamson–Hall analysis from the InN (0002)/(0004) peak widths (d) yields the microstrain $\varepsilon$ (e) and the interplanar spacing $d_{0002}$ (f) as functions of $\Delta\mu$ for the low-$\Delta\mu$ set.}

\end{figure*}

\section{Model and Equations}

The plasma-assisted MOCVD growth of InN is treated as a non-equilibrium surface reaction governed by the chemical potentials of the indium- and nitrogen-bearing gas species. The fundamental thermodynamic quantity controlling incorporation and defect formation is the gas--solid chemical-potential difference,
\begin{equation}
\Delta\mu \equiv \mu_{\mathrm{In}}(T,p_{\mathrm{In}})
+ \mu_{\mathrm{N}}(T,p_{\mathrm{N}})
- \mu_{\mathrm{InN}}^{(s)}(T),
\end{equation}
which quantifies the driving force for the reaction
\[
\mathrm{In}(g)+\mathrm{N}(g)\rightarrow\mathrm{InN}(s).
\]

For ideal gases, $\mu_i(T,p_i)=\mu_i^\circ(T)+k_BT\ln(p_i/p^\circ)$, yielding
\begin{equation}
\Delta\mu(T,p_{\mathrm{In}},p_{\mathrm{N}})
=
k_BT
\ln\!\left(
\frac{
p_{\mathrm{In}}p_{\mathrm{N}}
}{
p_{\mathrm{In},eq}p_{\mathrm{N},eq}
}
\right),
\label{eq:mu_ideal}
\end{equation}
where $(p_{\mathrm{In},eq},p_{\mathrm{N},eq})$ satisfy $\mu_{\mathrm{In}}+\mu_{\mathrm{N}}=\mu_{\mathrm{InN}}^{(s)}$ along the equilibrium gas–solid boundary.

In the experiments presented here, the true partial pressures of the activated nitrogen species are not directly measurable. Instead, an experimentally determined supersaturation parameter $\sigma$ is extracted from precursor supply and reactor conditions. This quantity is monotonic with the indium chemical activity, allowing $\Delta\mu$ to be represented experimentally as
\begin{equation}
\Delta\mu_{\mathrm{exp}}
=
k_B T \ln(1+\sigma),
\end{equation}
which rescales the data without altering sample ordering. All experimentally reported $\Delta\mu$ values correspond to this effective driving force.

\subsection*{Growth kinetics}

Incorporation arises from competition between forward insertion and reverse desorption events. The net incorporation flux is
\begin{equation}
J = J_f - J_b.
\end{equation}
Close to equilibrium, the flux–force relation follows Onsager’s linear law $J=L\Delta\mu$. However, over the experimentally explored range, where $\Delta\mu/k_BT$ is no longer small, the forward rate acquires an activated dependence,
\begin{equation}
J_f(\Delta\mu) \propto 
\exp\!\left(\frac{\Delta\mu}{E_g}\right),
\end{equation}
with $E_g$ an effective incorporation barrier. The macroscopic growth rate
\(
R=\Omega J
\)
then takes the form
\begin{equation}
R(\Delta\mu)
=
A\left[
\exp\!\left(\frac{\Delta\mu}{E_g}\right)-1
\right],
\label{eq:Rexp}
\end{equation}
where $A$ absorbs the attempt frequency and molecular volume. Equation~\eqref{eq:Rexp} reduces to the linear Onsager form for $\Delta\mu\!\ll\!k_BT$, but provides the correct exponential behavior observed experimentally over the full growth window.

\subsection*{Defect thermodynamics and Raman scattering}

Native point defects formed during growth are likewise governed by the chemical potentials. A defect $D$ created by adding or removing $n_\mathrm{In}$ indium atoms and $n_\mathrm{N}$ nitrogen atoms has formation free energy
\begin{equation}
\Delta G_{\mathrm{form}}(D)
=
\Delta G_0(D)
-
n_{\mathrm{In}}\mu_{\mathrm{In}}
-
n_{\mathrm{N}}\mu_{\mathrm{N}},
\end{equation}
which may be rewritten using $\mu_{\mathrm{In}}+\mu_{\mathrm{N}}=\mu_{\mathrm{InN}}^{(s)}+\Delta\mu$ as
\begin{equation}
\Delta G_{\mathrm{form}}(D)
=
\tilde G_D
-
\gamma_D\,\Delta\mu,
\qquad
\gamma_D=n_{\mathrm{In}}+n_{\mathrm{N}}.
\end{equation}
The defect concentration therefore follows
\begin{equation}
c_D(\Delta\mu)
=
c_D(0)\exp\!\left(
\frac{\gamma_D\Delta\mu}{k_BT}
\right).
\label{eq:defects}
\end{equation}

The non-polar Raman $E_2^{\mathrm{high}}$ phonon lifetime provides a direct probe of such point defects and local strain. Denoting the intrinsic lifetime by $\tau_{\mathrm{intr}}$ and the defect-induced contribution by a sum over defect species,
\begin{equation}
\frac{1}{\tau_{E2}(\Delta\mu)}
=
\frac{1}{\tau_{\mathrm{intr}}}
+
\sum_D A_D\,c_D(\Delta\mu),
\end{equation}
substituting Eq.~\eqref{eq:defects} gives the full exponential dependence
\begin{equation}
\frac{1}{\tau_{E2}(\Delta\mu)}
=
\frac{1}{\tau_{\mathrm{intr}}}
+
\sum_D A_D c_D(0)
\exp\!\left(
\frac{\gamma_D\Delta\mu}{k_BT}
\right).
\label{eq:tauE2full}
\end{equation}
Over very small ranges of $\Delta\mu$, this expression may be linearized; however, all analysis and fitting in this work employ the full exponential form~\eqref{eq:tauE2full} to remain consistent with experiment.

\begin{figure*}[t]
    \centering
    \includegraphics[width=0.99\textwidth]{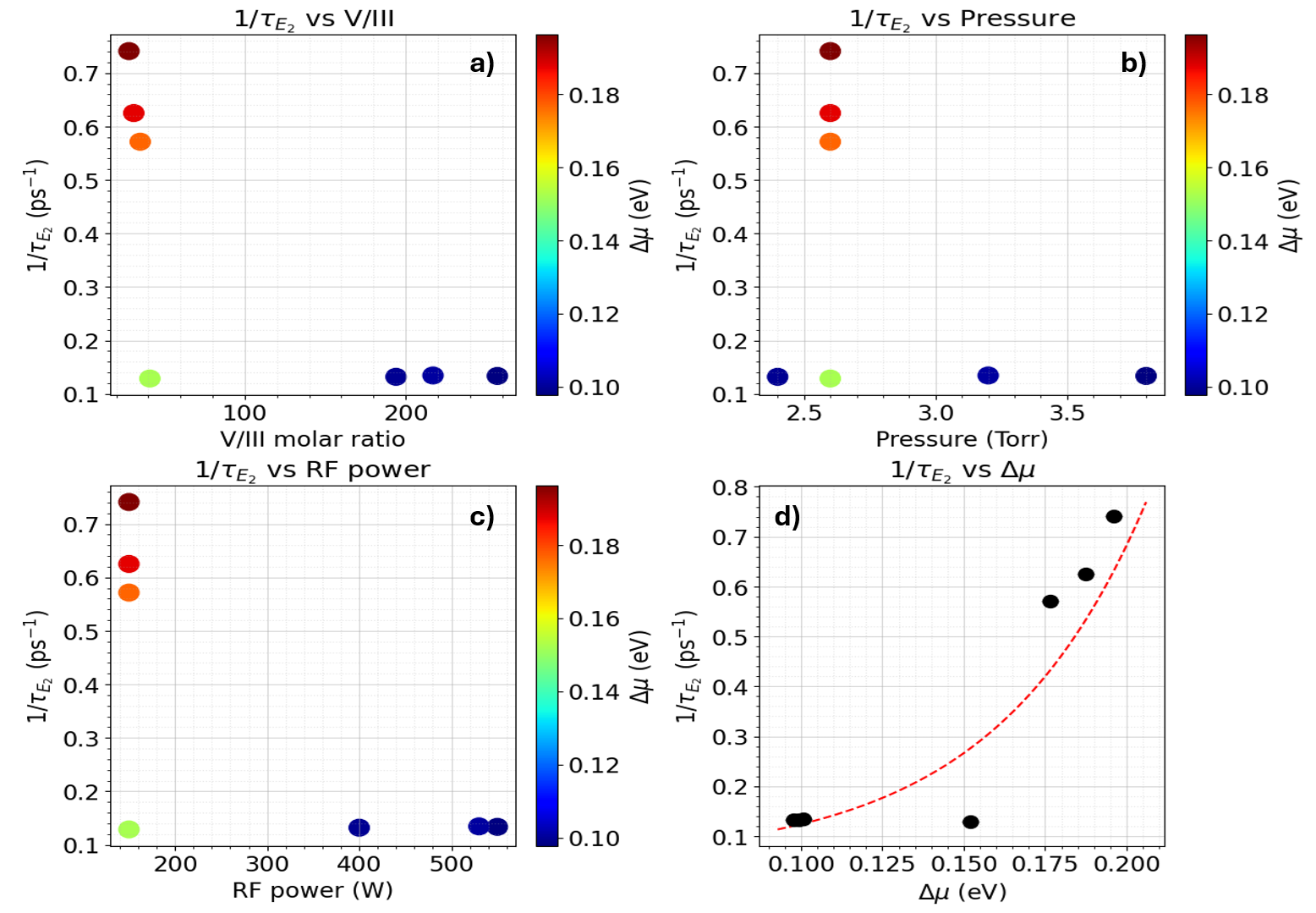}
    \\
    {\textbf{Figure 2:} Conventional process knobs do not organize defect scattering, while $\Delta\mu$ does. The $E_2^{\mathrm{high}}$ scattering rate $1/\tau_{E_2}$ (from Raman linewidths) plotted versus V/III ratio (a), reactor pressure (b), and RF plasma power (c) shows no unique monotonic dependence. When the same data are plotted versus the effective driving force $\Delta\mu$ (d), the dataset collapses onto a single activated trend (dashed line: fit). Points are color-coded by $\Delta\mu$.}

\end{figure*}

\subsection*{Nucleation and microstructure}

The grain-scale morphology is governed by the 2D nucleation free energy. A circular monolayer island of radius $r$ has free-energy cost
\begin{equation}
\Delta G(r)
=
2\pi r\,\gamma_{\mathrm{edge}}
-
\pi r^{2}\Delta\mu\,\Omega_{2D},
\end{equation}
where $\gamma_{\mathrm{edge}}$ is the step edge free energy and $\Omega_{2D}$ the surface area per InN formula unit. The critical radius follows from $\partial\Delta G/\partial r=0$,
\begin{equation}
r^\ast = \frac{\gamma_{\mathrm{edge}}}{\Delta\mu\,\Omega_{2D}},
\end{equation}
and the corresponding nucleation barrier is
\begin{equation}
\Delta G^\ast
=
\frac{\pi\gamma_{\mathrm{edge}}^{2}}{\Delta\mu\,\Omega_{2D}}.
\end{equation}
Thus the steady-state island (grain) density satisfies
\begin{equation}
N_g(\Delta\mu)
\propto
\exp\!\left[
-\frac{\Delta G^\ast}{k_B T}
\right]
=
\exp\!\left[
-\frac{\pi\gamma_{\mathrm{edge}}^{2}}{k_BT\,\Delta\mu\,\Omega_{2D}}
\right],
\end{equation}
implying the characteristic grain size
\begin{equation}
d(\Delta\mu)
\sim N_g^{-1/2}
\sim
\exp\!\left[
\frac{\pi\gamma_{\mathrm{edge}}^{2}}{2k_BT\,\Delta\mu\,\Omega_{2D}}
\right].
\end{equation}

Combined, Eqs.~\eqref{eq:Rexp}--\eqref{eq:tauE2full} and the nucleation relations demonstrate that the growth rate, defect population, Raman lifetimes, and microstructural length scales are all governed by the single thermodynamic driving force $\Delta\mu$. The coefficients $E_g$, $E_d$, $A$, $A_D$, $\gamma_D$, and $\gamma_{\mathrm{edge}}$ are extracted experimentally and discussed in the Results and Discussion section.

\begin{figure*}[t]
    \centering
    \includegraphics[width= 0.99\textwidth]{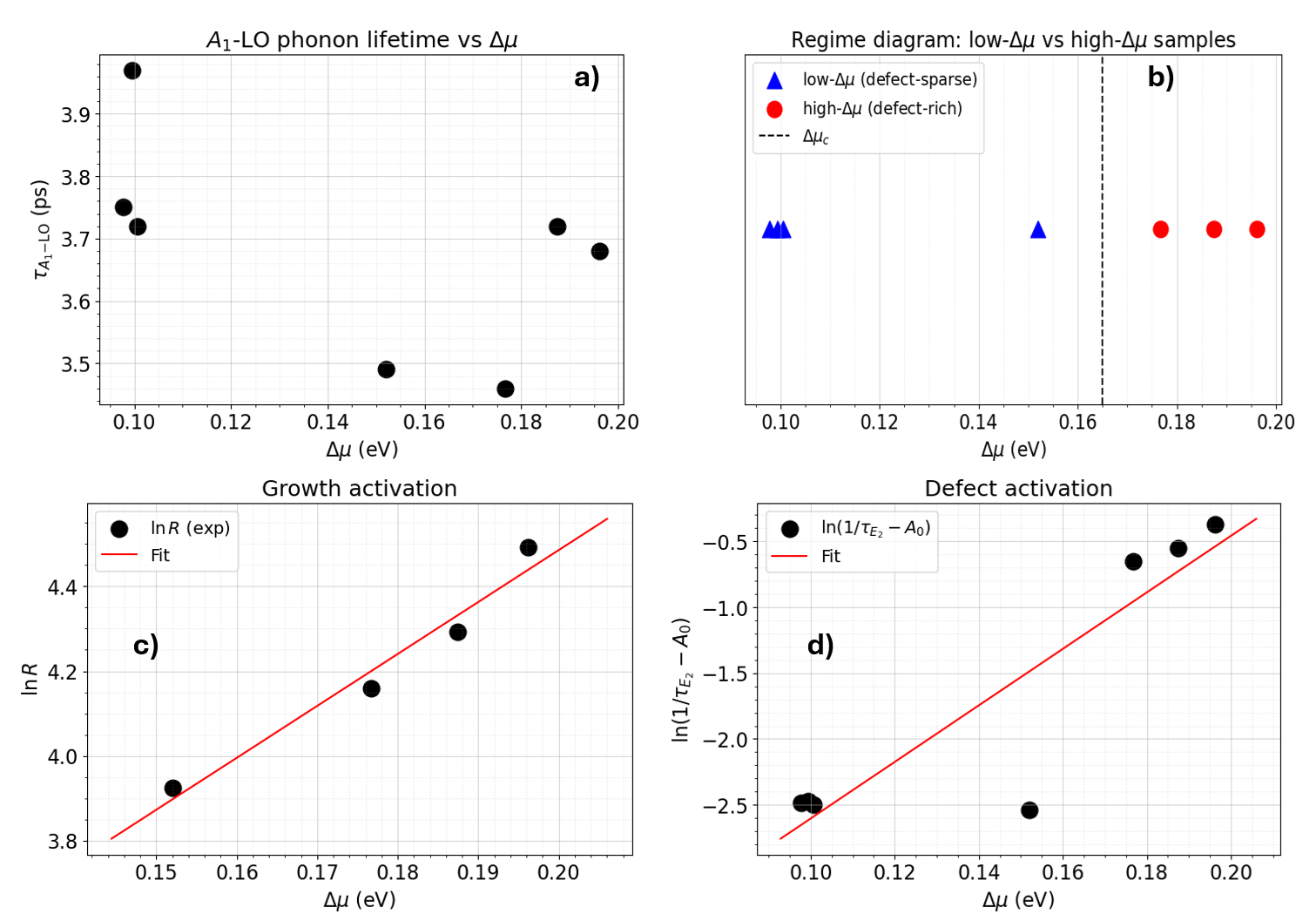}
    \\
    {\textbf{Figure 3:} Two-regime defect response and activated kinetics on the $\Delta\mu$ axis. The polar $A_1$–LO phonon lifetime shows minimal dependence on $\Delta\mu$ (a). A regime diagram separates low-$\Delta\mu$ (defect-sparse) and high-$\Delta\mu$ (defect-rich) samples with the crossover marked for reference (b). Log-linear representations of the growth rate, $\ln R$ vs $\Delta\mu$ (c), and the defect-mediated scattering contribution, $\ln(1/\tau_{E_2}-A_0)$ vs $\Delta\mu$ (d), yield the characteristic kinetic and defect-activation scales from the fitted slopes.}

\end{figure*}

\section{Kinetic Monte Carlo Simulation of Growth}

Kinetic Monte Carlo (KMC) simulations were performed to provide a microscopic realization of the $\Delta\mu$-dependent thermodynamic--kinetic framework and to test whether the experimentally observed exponential dependencies of growth rate and defect-mediated scattering can arise from local, thermally activated surface events. The KMC model implements the minimal set of incorporation, desorption, defect-creation, and defect-healing processes required to replicate the trends obtained experimentally.

The surface is represented by a two-dimensional hexagonal lattice containing
\[
N=L_x\times L_y
\]
sites with periodic lateral boundary conditions. Each site $i$ is assigned a discrete state
\[
s_i\in\{0,1,2\},
\]
where $s_i=0$ denotes an empty site, $s_i=1$ a correctly incorporated InN unit, and $s_i=2$ a locally distorted or defect-like configuration. The configuration $C=\{s_i\}$ is associated with an energy
\begin{equation}
E[C]
=
\sum_i \epsilon(s_i)
+
\sum_{\langle i,j\rangle} V(s_i,s_j),
\end{equation}
where $\epsilon(s_i)$ penalizes defect sites and $V(s_i,s_j)$ encodes a finite step-edge energy. This minimal energetic structure is sufficient to reproduce compact island formation and a nonzero nucleation barrier.

\begin{figure*}[t]
    \centering
    \includegraphics[width=0.99\textwidth]{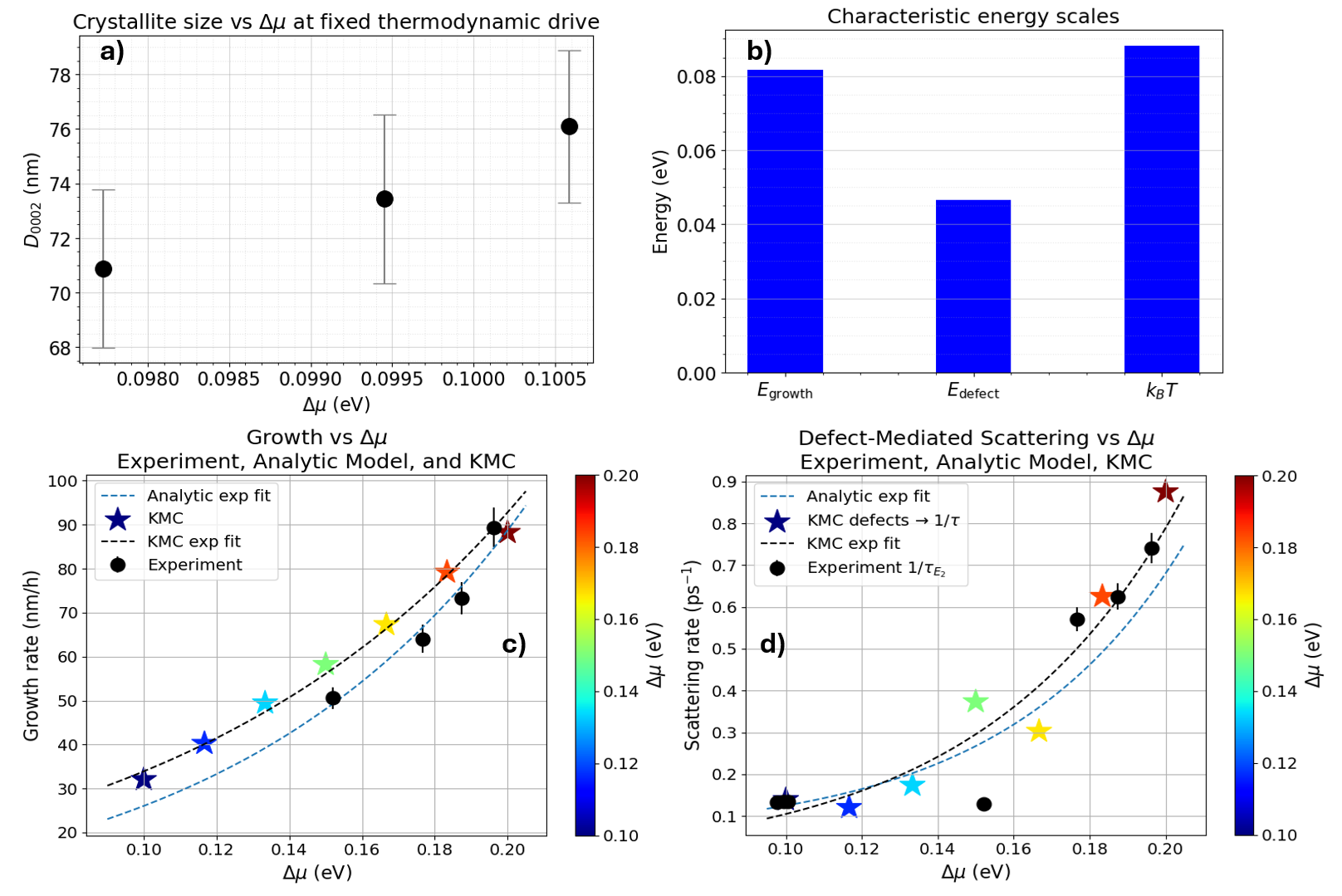}
    \\
    {\textbf{Figure 4:} Quantitative summary and model cross-check of the driving-force framework. For the low-$\Delta\mu$ subset, the out-of-plane coherence length $D_{0002}$ (Scherrer) remains essentially invariant within uncertainty at fixed driving force (a). The extracted characteristic scales for growth and defect activation are compared with $k_BT$ at the growth temperature (b). Growth rate vs $\Delta\mu$ (c) and defect-mediated scattering rate vs $\Delta\mu$ (d) show experiment (black circles), analytic model (blue dashed), and KMC results (stars; color indicates $\Delta\mu$) with corresponding exponential fits (black dashed).}

\end{figure*}

The dynamics consists of thermally activated events $C\to C_k'$ with rates
\begin{equation}
r_k(C;\Delta\mu)
=
\nu_k
\exp\!\left[
-\frac{E_k^\ddagger(C;\Delta\mu)}{k_BT}
\right],
\end{equation}
where $\nu_k$ is an attempt frequency and $E_k^\ddagger$ an activation barrier. Four event classes are included:

(i) incorporation of InN at empty sites or step edges,\\
(ii) desorption of weakly bound InN units,\\
(iii) creation of defect-like units, and\\
(iv) annihilation (healing) of such units.

The barriers for incorporation and desorption are directly coupled to the thermodynamic driving force. For incorporation events that increase the number of incorporated InN units,
\begin{equation}
E_{\mathrm{inc}}^\ddagger(C;\Delta\mu)
=
E_{\mathrm{inc},0}^\ddagger(C)
-
\alpha_{\mathrm{inc}}\Delta\mu,
\end{equation}
and for desorption events,
\begin{equation}
E_{\mathrm{des}}^\ddagger(C;\Delta\mu)
=
E_{\mathrm{des},0}^\ddagger(C)
+
(1-\alpha_{\mathrm{inc}})\Delta\mu,
\end{equation}
with $0\le\alpha_{\mathrm{inc}}\le1$. This construction ensures detailed balance in the limit $\Delta\mu\to 0$, and produces the exponential enhancement of incorporation with increasing $\Delta\mu$ consistent with the growth law extracted experimentally.

Defect creation and annihilation are treated analogously. Writing the formation free energy of a defect-like local configuration as
\[
\Delta G_{\mathrm{form}}(D)
=
\tilde G_D - \gamma_D \Delta\mu,
\]
the corresponding activation barriers are 
\begin{equation}
E_{\mathrm{def},c}^\ddagger(C;\Delta\mu)
=
E_{\mathrm{def},c,0}^\ddagger(C)
-
\beta_c\,\gamma_D\,\Delta\mu,
\end{equation}
and
\begin{equation}
E_{\mathrm{def},a}^\ddagger(C;\Delta\mu)
=
E_{\mathrm{def},a,0}^\ddagger(C)
+
(1-\beta_c)\,\gamma_D\,\Delta\mu,
\end{equation}
with $0\le\beta_c\le1$. In steady state, this yields the defect concentration
\begin{equation}
c_D^{\mathrm{KMC}}(\Delta\mu)
\propto
\exp\!\left(
\frac{\gamma_D\Delta\mu}{k_BT}
\right),
\end{equation}
consistent with the exponential dependence observed in the Raman $E_2^{\mathrm{high}}$ scattering rate.

The stochastic evolution is implemented using a rejection-free Gillespie algorithm. All possible events $\{k\}$ and their rates $r_k(C;\Delta\mu)$ are enumerated, and the total rate
\begin{equation}
R_{\mathrm{tot}}(C;\Delta\mu)
=
\sum_k r_k(C;\Delta\mu)
\end{equation}
is computed. An event $k$ is selected with probability
\[
P_k=\frac{r_k(C;\Delta\mu)}{R_{\mathrm{tot}}(C;\Delta\mu)},
\]
and the time is advanced by
\[
\Delta t
=
-\frac{\ln u}{R_{\mathrm{tot}}(C;\Delta\mu)},
\]
with $u$ a uniform random number in $(0,1)$. The updated configuration $C_k'$ becomes the state for the next iteration, and the procedure is repeated until a steady state is reached.

From the KMC trajectories, observables are extracted for direct comparison with experiment. The net incorporation flux gives the simulated growth rate
\begin{equation}
R_{\mathrm{KMC}}(\Delta\mu)
=
\frac{\Delta N_{\mathrm{InN}}\,\Omega}{\Delta t_{\mathrm{sim}}},
\end{equation}
where $\Delta N_{\mathrm{InN}}$ is the change in incorporated units over the sampling interval. The steady-state defect concentration is
\begin{equation}
c_D^{\mathrm{KMC}}(\Delta\mu)
=
\frac{N_D}{N_{\mathrm{InN}}},
\end{equation}
with $N_D$ and $N_{\mathrm{InN}}$ the numbers of defect-like and incorporated sites in the sampling region. A simulated ``Raman'' scattering rate is defined analogously to experiment:
\begin{equation}
\frac{1}{\tau_{E2}^{\mathrm{KMC}}(\Delta\mu)}
=
\frac{1}{\tau_{\mathrm{intr}}}
+
\sum_D A_D\,c_D^{\mathrm{KMC}}(\Delta\mu).
\end{equation}
Finally, the in-plane island density $N_g^{\mathrm{KMC}}$ is obtained from cluster analysis, and the corresponding grain size is estimated as
\begin{equation}
d_{\mathrm{KMC}}(\Delta\mu)
\sim
\left[N_g^{\mathrm{KMC}}(\Delta\mu)\right]^{-1/2},
\end{equation}
allowing direct comparison with the crystallite size extracted from XRD.

\section{RESULTS AND DISCUSSION}

The central experimental result of this work is that the many nominal growth knobs in PA-MOCVD InN can be expressed through a single effective coordinate, the thermodynamic driving force $\Delta\mu$. The dataset covers large changes in V/III ratio, reactor pressure and RF plasma power. When the growth rate and the defect sensitive Raman response are plotted against $\Delta\mu$ they become smooth and systematic. When the same quantities are plotted against any one reactor setting they remain non universal. The goal of this section is therefore not to catalog growth recipes again. Instead we show how one driving force axis organizes three things at once. These are how fast the film grows, where defect mediated scattering switches on strongly, and which aspects of structural coherence stay nearly unchanged inside the defect sparse window.

The first internal check comes from the Raman spectra. Representative spectra in Fig.~1(a,b) show the expected InN phonon modes together with sapphire lines. They also show that the defect sensitive mode and the control mode respond differently when we move along the same driving force axis. In this work the nonpolar $E_{2}^{\mathrm{high}}$ mode is used as the main probe of localized lattice disorder. Its linewidth derived lifetime, or equivalently its scattering rate, responds strongly to point defects and local strain fields that break translational symmetry without requiring long range polar fields. The polar $A_{1}$--LO mode is less sensitive to such local nonpolar disorder and is strongly shaped by long range polar coupling. The fact that the $A_{1}$--LO lifetime is essentially invariant with $\Delta\mu$ (Fig.~3a) therefore acts as an internal control. The large changes seen in the $E_{2}^{\mathrm{high}}$ lifetime are not a generic broadening of everything. They are consistent with an extra short range defect mediated scattering channel that selectively degrades the nonpolar phonon coherence.

With that spectroscopic baseline in place the main question is whether the defect mediated scattering can be written as a single valued function of one physically meaningful coordinate. When $1/\tau_{E_{2}}$ is plotted against V/III ratio, pressure or RF power no unique monotonic trend appears. Samples with similar scattering rates occur at very different values of each knob (Fig.~2a--c). When the same data are plotted versus $\Delta\mu$ they fall on a single activated curve (Fig.~2d). This collapse is the practical evidence that within this growth window $\Delta\mu$ captures the net incorporation bias relevant for defect activation more faithfully than any single reactor parameter. In more concrete terms the plot shows that V/III ratio, pressure and RF power are not independent quality directions. They act together through their combined effect on the effective driving force.

The collapsed curve is not just a gradual variation. It breaks naturally into two regimes on the same $\Delta\mu$ axis. At low driving force the scattering rate $1/\tau_{E_{2}}$ stays small and the $E_{2}^{\mathrm{high}}$ lifetime is long. This defines a defect sparse regime in the present metric. At higher driving force the scattering rate rises rapidly and the lifetime collapses. This defines a defect rich regime. Because the dataset is discrete we do not claim a sharply defined critical point. Instead we bracket the onset between the two nearest measured values. We label the crossover by the midpoint and the bracket width
\begin{equation}
\Delta\mu_{\times} \approx 0.165 \pm 0.015~\mathrm{eV},
\end{equation}
(30)
which lies between the measurements at $\Delta\mu = 0.152~\mathrm{eV}$ and $\Delta\mu = 0.177~\mathrm{eV}$. The regime diagram in Fig.~3b is a visual summary of this separation. The quantitative statement is that the crossover takes place somewhere between those two points, not that the system has a unique singular threshold.

Once everything is expressed in terms of $\Delta\mu$ the rise of defect mediated scattering is well described by an activated law. In Fig.~3d we subtract a baseline $A_{0}$ and plot $\ln(1/\tau_{E_{2}} - A_{0})$ versus $\Delta\mu$. The result is nearly linear and is consistent with
\begin{equation}
\left(\frac{1}{\tau_{E_{2}}} - A_{0}\right) \propto \exp\!\left(\frac{\Delta\mu}{E_{d}}\right),
\end{equation}
(31)
with an activation scale $E_{d}$ of order 0.05~eV, summarized with other scales in Fig.~4b. This number is not used as a purely empirical fit. In the defect bias picture that underlies both the experiment and the KMC model we interpret it by matching the fitted activated form to a driving force biased defect propensity,
\begin{equation}
\exp\!\left(\frac{\beta_{\mathrm{def}} \gamma_{D} \Delta\mu}{k_{\mathrm{B}} T}\right)
\equiv
\exp\!\left(\frac{\Delta\mu}{E_{d}}\right)
\;\Rightarrow\;
E_{d} \approx \frac{k_{\mathrm{B}} T}{\beta_{\mathrm{def}} \gamma_{D}}
\end{equation}
(32)
At the fixed growth temperature $T = 1048~\mathrm{K}$ we have $k_{\mathrm{B}} T \simeq 0.090~\mathrm{eV}$. With $E_{d} \simeq 0.05~\mathrm{eV}$ this gives $\beta_{\mathrm{def}} \gamma_{D} \simeq 1.8$. This is naturally consistent with an effective $\gamma_{D}$ near 2, which corresponds to a defect process that involves exchange of roughly two atoms with the lattice along the activated pathway, and a bias factor $\beta_{\mathrm{def}}$ in the range 0.9 to 1 that measures how efficiently the driving force tilts that barrier. In this way the experimental $E_{d}$ connects directly to a physically meaningful bias strength in the same language used for the minimal KMC model.

The same driving force coordinate that organizes disorder also organizes incorporation kinetics. When we plot the growth rate as $\ln R$ versus $\Delta\mu$ the data lie very close to a straight line (Fig.~3c). This is consistent with an activated incorporation law of the form
\begin{equation}
R(\Delta\mu) \approx A \left[ \exp\!\left(\frac{\Delta\mu}{E_{g}}\right) - 1 \right]
\end{equation}
(or, away from the near-equilibrium
(33)
The extracted kinetic scale $E_{g}$ is about 0.08~eV (Fig.~4b), which is comparable to $k_{\mathrm{B}} T$ at 1048~K. This near equality is a compact way to say that growth in this PA-MOCVD window is only weakly driven in the thermodynamic sense. Modest changes in $\Delta\mu$ produce exponentially large changes in incorporation flux. The same idea is visible when we overlay the experimental points, the analytic activated form and the KMC results on one $\Delta\mu$ axis (Fig.~4c). The functional dependence is reproduced without having to redefine anything in terms of reactor specific knobs.

Structural measures give a complementary view of what remains nearly unchanged within the defect sparse window. XRD spectra for the low $\Delta\mu$ subset (Fig.~1c) show InN (0002) and (0004) reflections alongside strong sapphire peaks. For samples with closely matched driving force the InN peak shapes remain similar. Using these two reflections in a Williamson--Hall construction (Fig.~1d) we estimate the microstrain $\varepsilon$ and separate strain and size contributions in the standard way. Within the low $\Delta\mu$ subset the microstrain and the $d_{0002}$ spacing change only modestly over the narrow interval in driving force (Fig.~1e,f). This is consistent with the picture that the defect sparse regime corresponds to a relatively stable local lattice environment in the sense probed by Raman. Independently the out of plane coherence length $D_{0002}$ extracted from the symmetric (0002) reflection stays essentially constant within uncertainty across the same low $\Delta\mu$ subset (Fig.~4a) even though the underlying reactor settings differ strongly between those samples. Because symmetric XRD is mainly sensitive to out of plane coherence rather than in plane grain size we interpret this conservatively. 


Finally the mechanistic role of $\Delta\mu$ is supported by the minimal kinetic Monte Carlo picture in which both incorporation and defect activation propensities are biased by the same driving force coordinate. The intent is not to claim that KMC provides absolute time scales in a plasma chemistry environment that it does not explicitly resolve. The point is that a small set of $\Delta\mu$ biased activated events is enough to reproduce the functional forms seen in experiment. When KMC outputs are placed on the same $\Delta\mu$ axis as the data the growth trend and the defect mediated scattering trend show the same activated behavior (Fig.~4c,d). The characteristic experimental scales $E_{g}$ and $E_{d}$ are naturally of the same order as the effective bias parameters used in the minimal model. Together with the $A_{1}$--LO control this closes the loop. One driving force coordinate organizes incorporation, switches on a nonpolar defect scattering channel and leaves out of plane coherence nearly unchanged within the defect sparse window of this dataset.

Our results show that within this growth window the apparent complexity of PA-MOCVD InN is not an irreducible dependence on several reactor knobs. Instead it can be expressed compactly in terms of one effective driving force coordinate. That coordinate collapses the growth kinetics onto an activated law with $E_{g}$ of order $k_{\mathrm{B}} T$ and it organizes a crossover to defect rich growth in which the $E_{2}^{\mathrm{high}}$ lifetime collapses and the defect mediated scattering rate becomes exponentially activated with a scale $E_{d}$ near 0.05~eV. The combined Raman, XRD and KMC evidence supports a consistent physical picture. As $\Delta\mu$ increases the incorporation rate accelerates and defect activation pathways become biased, and this produces a clear change in nonpolar phonon coherence that is not mirrored by the polar $A_{1}$--LO control mode.

\section{Conclusions}
We found that plasma assisted MOCVD growth of InN can be understood in terms of a single effective thermodynamic driving force $\Delta\mu$ built from an experimentally defined supersaturation index. Once the data are expressed on this axis the growth rate follows an activated law with $E_{g}$ close to 0.08~eV, the defect sensitive Raman metric shows a clear crossover from defect sparse to defect rich growth between $\Delta\mu = 0.152~\mathrm{eV}$ and $0.177~\mathrm{eV}$ with an activated increase set by $E_{d}$ around 0.05~eV, and the out of plane structural coherence stays essentially unchanged for samples that share the same driving force in the low $\Delta\mu$ window. In other words the complicated dependence on V over III ratio, pressure and RF power reduces to a simple dependence on how hard the system is driven thermodynamically. The kinetic Monte Carlo simulations recover the same exponential trends and the same separation into defect sparse and defect rich regimes that we see in experiment. The extracted experimental scales $E_{g}$ and $E_{d}$ fall naturally in the range of the effective bias parameters needed in the model. Hence the KMC loses the loop on the experimental analysis that a single driving force coordinate is enough to control incorporation, trigger a nonpolar defect scattering channel and leave out of plane coherence largely unchanged over the relevant growth window.


\bibliography{references}

\bibliographystyle{unsrtnat}

\end{document}